\documentclass{ragtime} 
\usepackage{times}
\title[]{Flux ropes in SANE disks}

\author[M. \v{C}emelji\'{c}, 
        F. Yuan    
        and H. Yang]
       {Miljenko \v{C}emelji\'{c}\at{1,2,3,a} 
        Feng Yuan\at[]{1} 
        and Hai Yang\at[]{1}\\
        \ins{1}Shanghai Astronomical Observatory, Chinese Academy of        Sciences,\splitins[1]
        80 Nandan Road, Shanghai 200030, China\\
        \ins{2}Nicolaus Copernicus Astronomical Center, Polish Academy of 
        Sciences,\splitins[2]
        Bartycka 18, 00-716 Warsaw, Poland\\
        \ins{3}Academia Sinica, Institute of Astronomy and Astrophysics,\splitins[2]
        P.O. Box 23-141, Taipei 106, Taiwan\\
        \ins{a}\Email{miki@camk.edu.pl}} 

\RequirePackage{ifpdf}
\usepackage{hyperref}

\graphicspath{{cem/}}

\begin{document}

\begin{abstract}
Three-dimensional numerical simulations of a hot accretion flow around a
supermassive black hole are performed using the general relativity magneto-hydrodynamic (GRMHD) code Athena++. We focus on the case of SANE,
with the initial magnetic field consisting of multiple loops with oppositely directed poloidal magnetic field
in the torus. Using the simulation data, we investigate the formation of flux ropes, follow the forming of flux ropes atop
the disk, and their release into corona.
\end{abstract}


\section{Introduction}\label{intro}

In the accreting systems, large scale jets are usually steady, while
episodic jets are sometimes related to flares, which are observed on the
smaller scale. One such example is Sgr~A*, a massive black hole in the Galactic
centre, where we observe radio, infrared and X-ray flares several times a
day. It was concluded that delays in peaks in the light curves at different
wavebands and their fast rise and slow decay in the brightness and
polarisation are related to the ejection and expansion of plasmoids from the
accretion flow. Knots in the jets are also observed, e.g. in 3C 120 and
M87, and could be related to episodic emission. There are models, like e.g. 
\cite{BZ77} and \cite{BP82} for continuous jets,
but we still do not have a viable model for episodic jets. In \cite{YuanFengetal09},
such a model was proposed, in analogy with Coronal Mass Ejections
(CMEs) in the Sun, with the closed magnetic field lines emerging from the
main body of the accretion flow, expelled to the corona region: The
foot-points of the magnetic loops are positioned in the turbulent accretion
flow, and their twisting results in magnetic reconnection, forming the flux
ropes. Because of the ongoing reconnection below such a flux rope, the
magnetic tension force weakens, and the initial equilibrium between the
magnetic tension and the magnetic pressure is not maintained. The flux ropes
will be accelerated outwards, forming the episodic jet. The flares,
observed from such jets, are from the emission originating from the
electrons accelerated by the reconnection. In \cite{Shende19}, another model was
proposed, in analogy with Toroidal Instability from tokamak research and also used to model the CMEs.

\citet{Nathanail20} present results of two-dimensional (2D) GRMHD simulations with the Black Hole Accretion Code (BHAC, \citet{Porthetal19}), with Adaptive Mesh Refinement (AMR) of both Magnetically Arrested and Standard and Normal Evolution (MAD and SANE) discs. Different initial magnetic field configurations and resolutions are chosen. They find the formation of copious plasmoids and describe their outward motion.  Similar simulations based on the same code, but with the physical resistivity included, are presented in \citet{Ripperda20}. They show no difference in results between the ideal and weakly resistive simulations. They conclude that 2D ideal MHD simulations, with only the numerical resistivity dissipating the magnetic field, can capture the physics.

In this work, we perform 3D GRMHD simulations to investigate the formation of magnetic flux ropes, checking the scenario proposed by \cite{YuanFengetal09}. 
\section{Numerical simulations setup}\label{numsetup}
\begin{figure}[t]
\begin{center}
\includegraphics[width=0.35\linewidth,height=.35\linewidth]{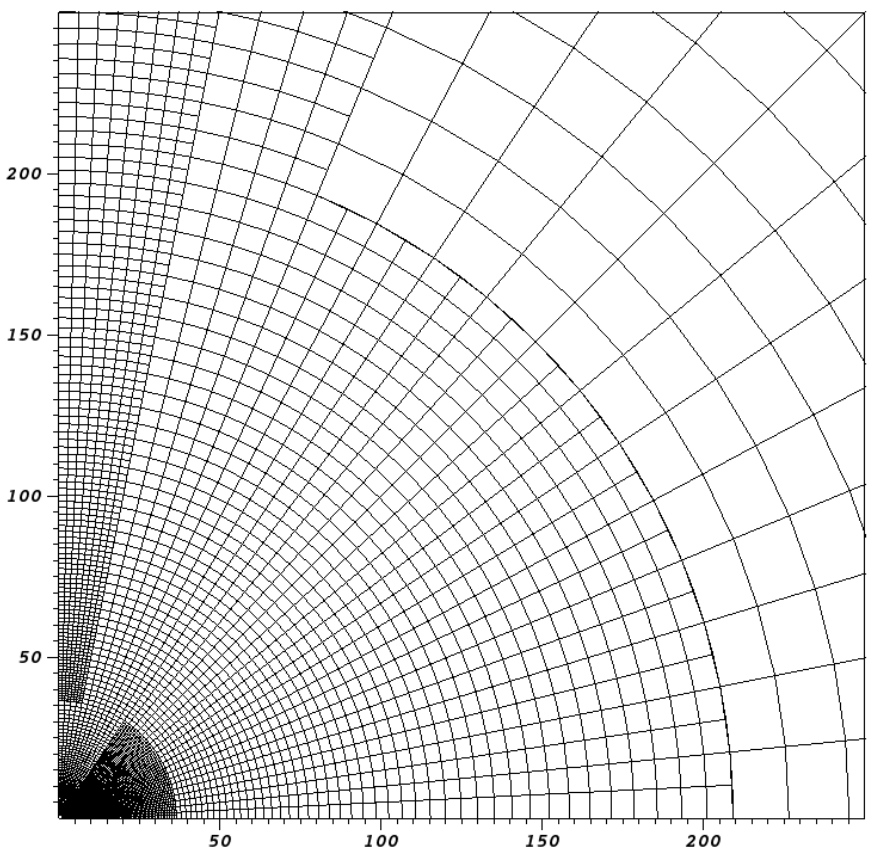}
\includegraphics[width=0.35\linewidth,height=.35\linewidth]{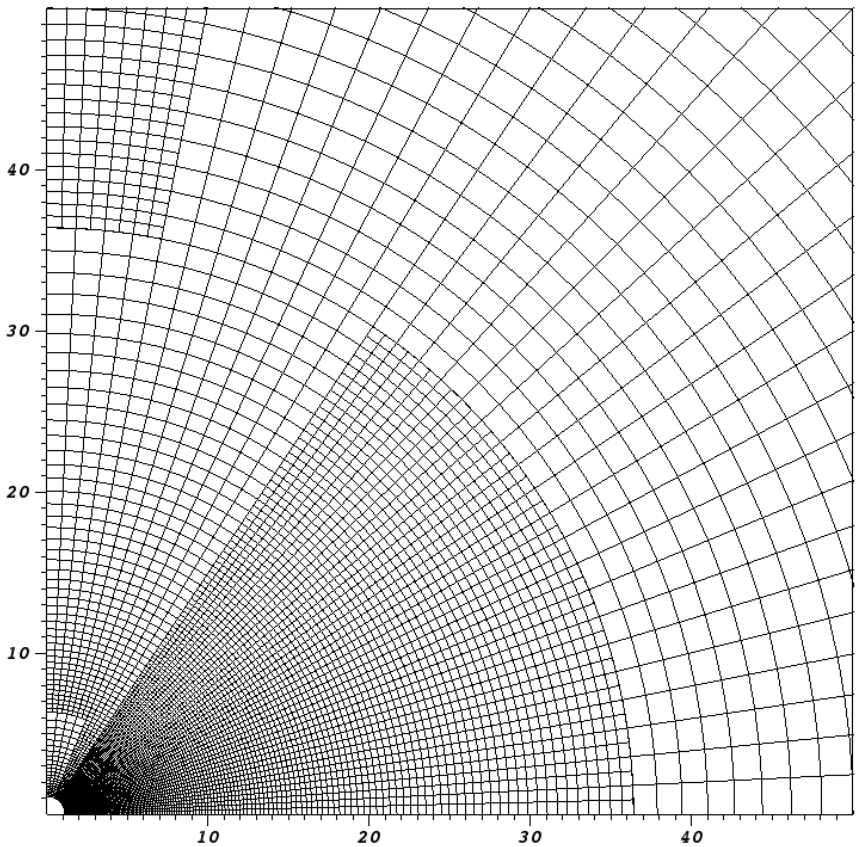}
\end{center}
\caption{We use static mesh refinement for the grid, to obtain largest resolution where it is most needed. Resolution is $R\times\theta\times\varphi=(288\times 128\times 64)$ grid cells in spherical coordinates, in a physical domain reaching to 1200 gravitational radii. The different refinements used in this grid are shown.}
\label{grid}
\end{figure}
\begin{figure}[t]
\begin{center}
\includegraphics[width=0.31\linewidth,height=.5\linewidth]{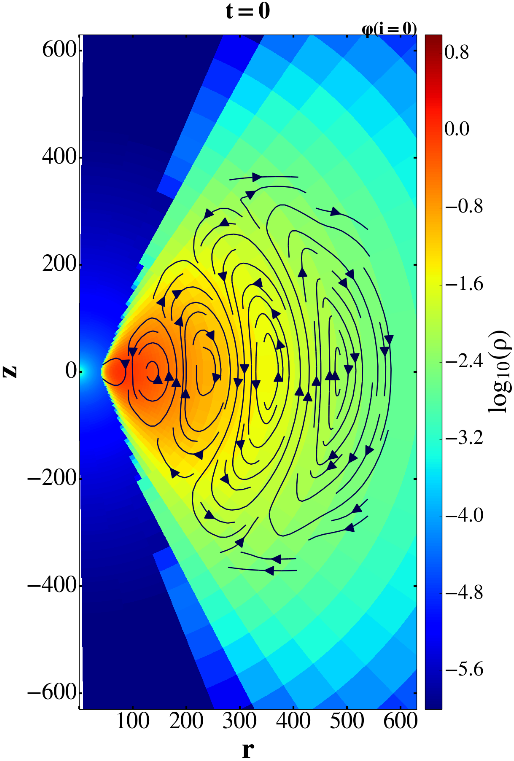}
\includegraphics[width=0.31\linewidth,height=.4\linewidth]{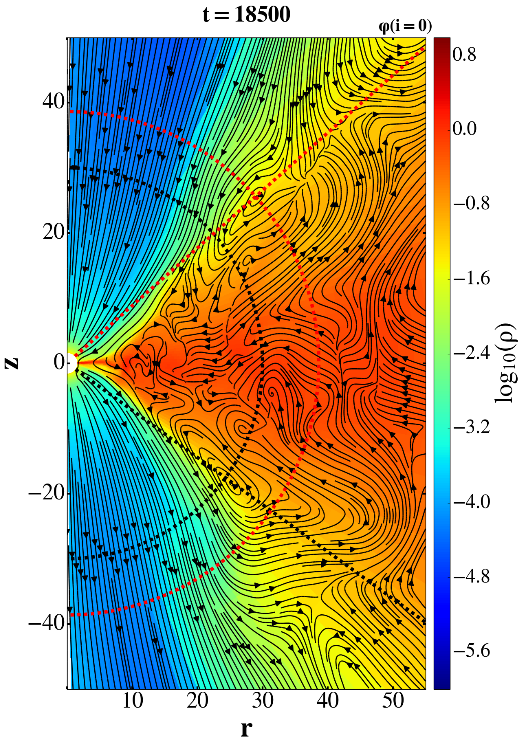}
\includegraphics[width=0.31\linewidth,height=.4\linewidth]{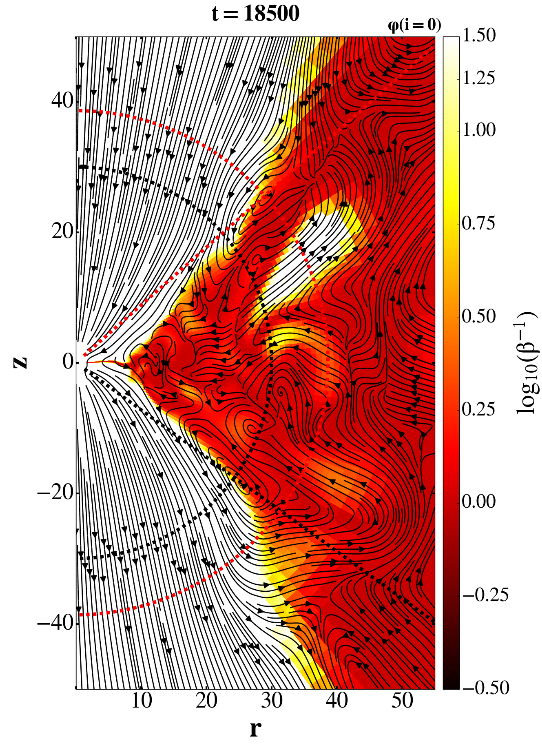}
\end{center}
\caption{Left panel: The initial setup in our SANE simulation. The color denotes the density,  solid lines denote the poloidal magnetic field with arrows showing the field direction. Middle panel: A zoomed-in snapshot in the result after $t=18500~r_{\mathrm g}/c$. The positions of two magnetic islands are marked with the cross of two dotted black lines and two dotted red
lines, respectively. Right panel: Same with the middle panel, but the color shows  the plasma $\beta=P_{\rm gas}/P_{\rm mag}$. The two magnetic islands are located  at the surface with plasma $\beta\sim 1$.}
\label{init}
\end{figure}
We perform numerical simulations using the GRMHD code Athena++ \citep{Whiteetal16} in full 3D, solving the ideal MHD equations in the Kerr metrics, in Kerr-Schild (horizon penetrating) coordinates. Resolution is $R\times\theta\times\varphi=(288\times 128\times 64)$ grid cells in spherical coordinates, in a physical domain reaching to 1200 gravitational radii, $r_g=GM/c^2$. We use different refinements in this grid, as shown in Fig.~\ref{grid}. The staggered mesh Constrained Transport (CT) method is applied to maintain the divergence-free magnetic field. Static mesh refinement is used for the grid, to obtain largest resolution where it is most needed. Initial configuration of density and magnetic field in our SANE simulation is shown in Fig.~\ref{init}. The central object is not rotating.

\section{Formation and motion of the flux rope}\label{fluxrope}
\begin{figure}[t]
\begin{center}
\includegraphics[width=0.4\linewidth,height=.37\linewidth]{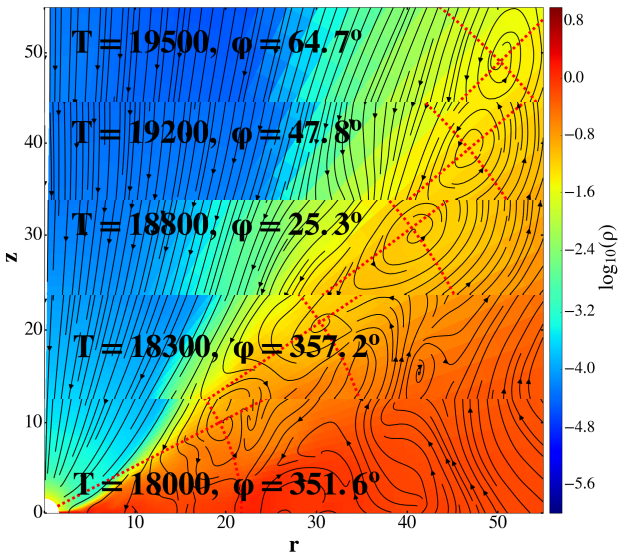}
\includegraphics[width=0.4\linewidth,height=.37\linewidth]{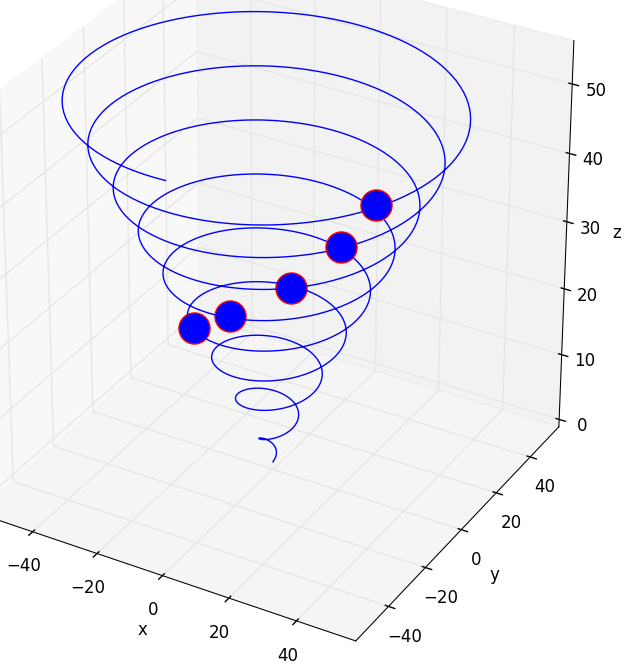}
\end{center}
\caption{Left panel: Outward motion of the magnetic island in our simulation, in different colatitudinal planes. Right panel: Spiralling-out of the magnetic island in a schematic plot of the trajectory in 3D. Positions of the blue circles are chosen to approximately represent magnetic islands from the left panel.}
\label{spir}
\end{figure}
\begin{figure}[t]
\begin{center}
\includegraphics[width=0.32\linewidth,height=.25\linewidth]{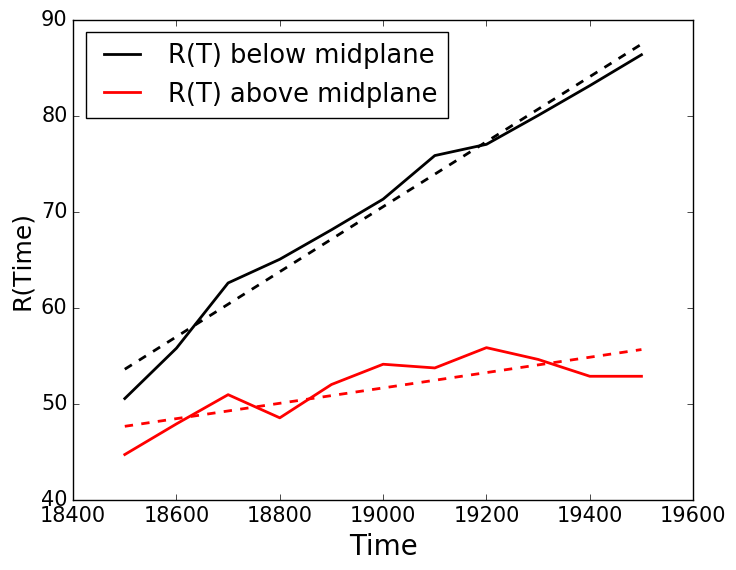}
\includegraphics[width=0.32\linewidth,height=.25\linewidth]{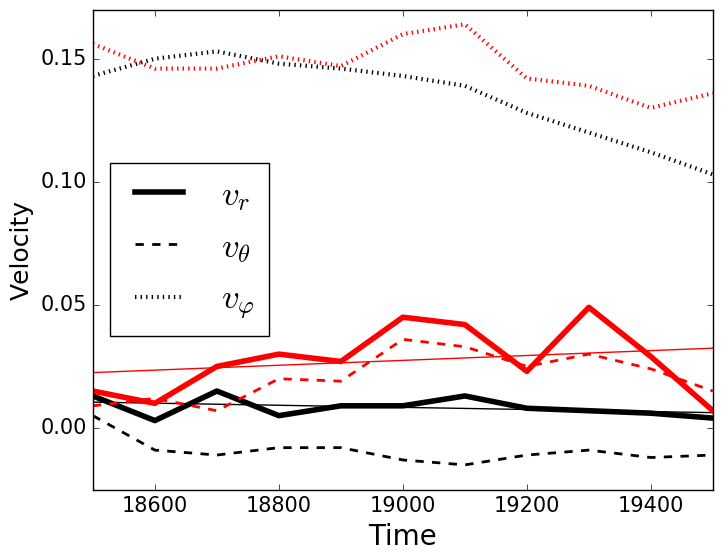}
\includegraphics[width=0.32\linewidth,height=.26\linewidth]{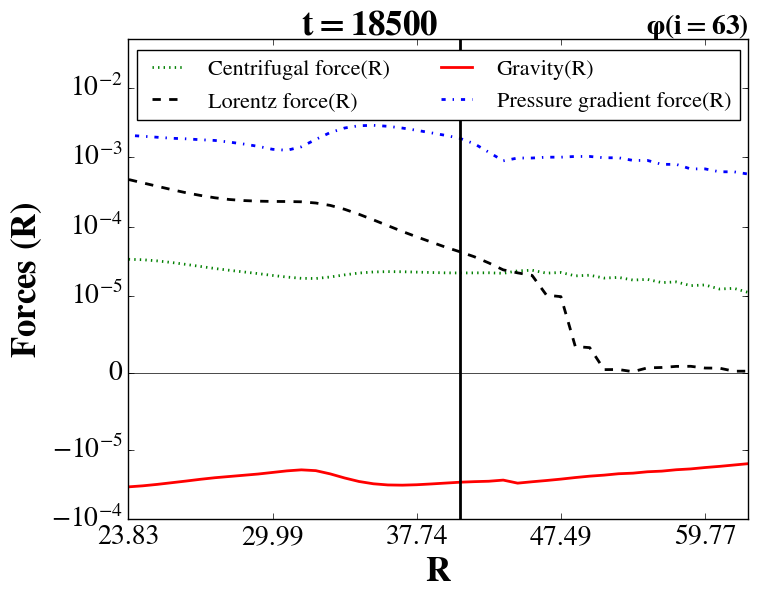}
\end{center}
\caption{Time evolution of the position of the center of magnetic island (left panel) and velocity of the material near the center of magnetic island (middle panel) for two magnetic islands from Fig.~\ref{init}. Slopes of the least square fits shown in dashed lines in the left panel are 0.03~c and 0.01~c for the black and red lines, respectively. In the middle panel, least square fits of the radial velocity components are shown by the corresponding color thin solid lines. In the right panel are shown forces, along the radial direction, on the material in the flux rope above the disk mid-plane, near the center of the magnetic island at $t=18500~r_{\mathrm g}/c$.}
\label{posvel}
\end{figure}
\begin{figure}[t]
\begin{center}
\includegraphics[width=0.48\linewidth,height=.7\linewidth]{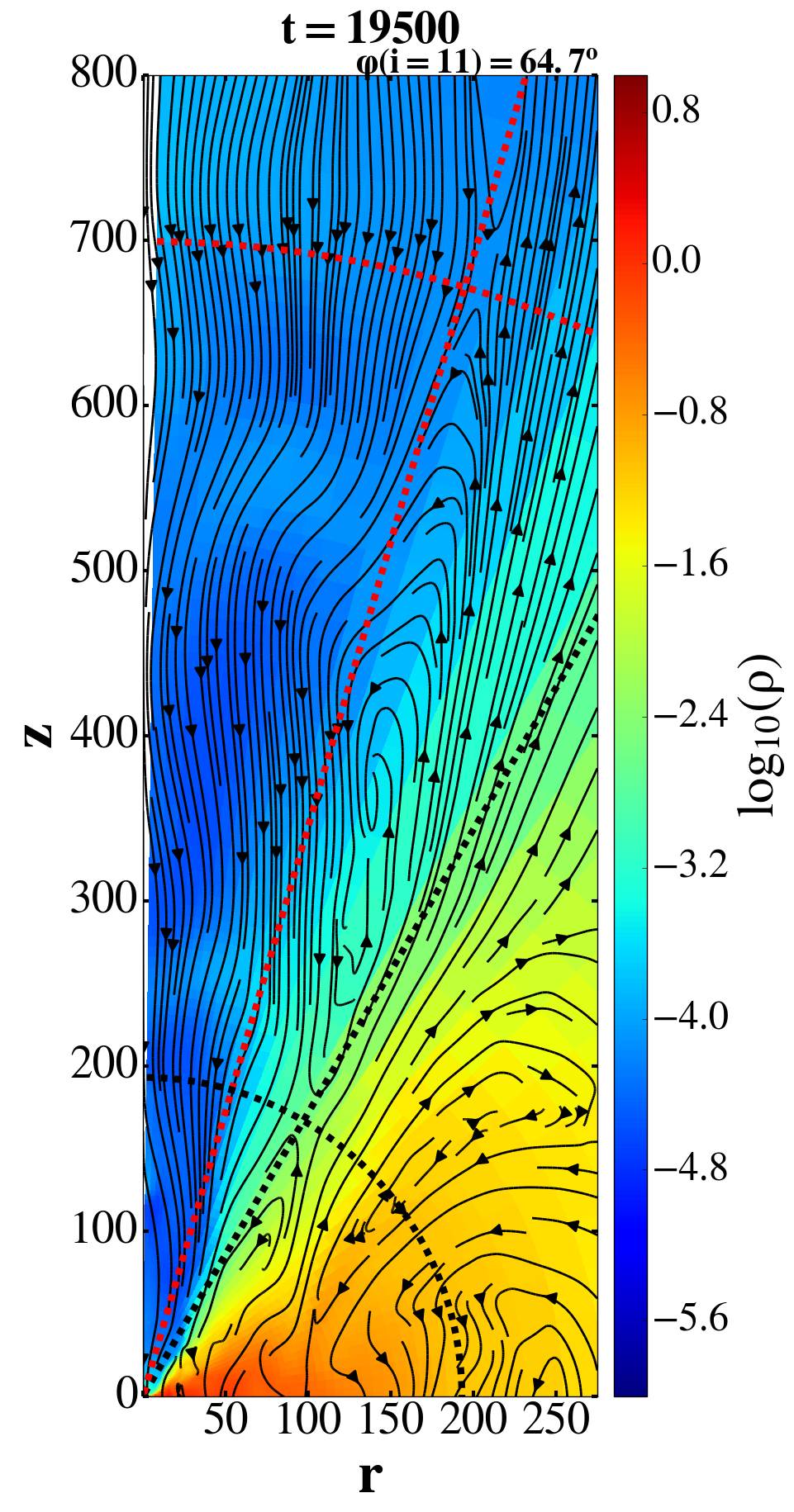}
\vspace*{-9cm}\includegraphics[width=0.5\linewidth,height=.35\linewidth]{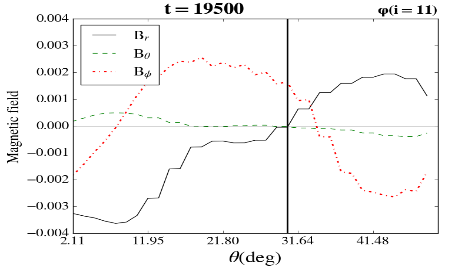}\\
\hspace*{6cm}\vspace*{-1cm}\includegraphics[width=0.5\linewidth,height=.35\linewidth]{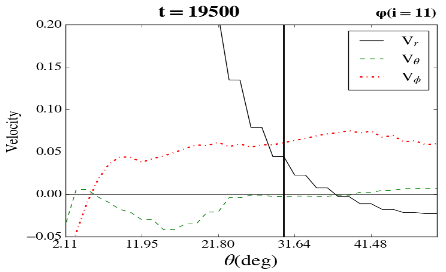}
\vspace*{5.5cm}
\end{center}
\caption{Two reconnection layers with a magnetic island between them in a snapshot at $t=19500~r_{\mathrm g}/c$ in our simulation are marked with the black and red dotted lines in the left panel. In the right panels are shown the velocity and magnetic field components at the same time, along a part of the black dotted line circle passing through the reconnection layer. A signature of reconnection, change in the direction of poloidal velocity and all three components of magnetic field, is visible in the projections in $\theta$-direction. A similar signature is obtained in the reconnection layer positioned at the intersection of red dotted lines. }
\label{velbs}
\end{figure}
Interchanging directions of the initial magnetic field in the torus prevent the field to grow too large, and magneto-rotational instability can provide the dissipation for successful accretion of material towards the central object. We perform our simulation until $t=40000~r_{\mathrm g}/c$. A snapshot at $t=18500~r_{\mathrm g}/c$ during the evolution of the accretion flow is shown in Fig.~\ref{init}. We find magnetic islands in the colatitudinal $(R, \theta)$ planes, at different azimuths $\varphi$, after the relaxation from the initial conditions and stabilization of the flows. Such magnetic islands start forming after about $t=15000~r_{\mathrm g}/c$. They periodically emerge from the disk surface at similar radii (azimuthal angle $\varphi$ changes with the rotation of the disk), with period of about $t=1000~r_{\mathrm g}/c$. The magnetic islands are extended in the azimuthal direction, forming magnetic flux ropes of various lengths. We trace the extension of the flux ropes in $\varphi$ direction, which is typically about $120^\circ$ or less, and perform slices in the middle of their length at different times--as shown in Fig.~\ref{spir}. In the same Figure we give a sketch in 3D of the counter-clockwise spiral trajectory of the indicated flux rope cross-sections. To understand the launching and motion of the flux rope, we measure positions of the magnetic islands and velocity of the material near their centers in time--see Fig.~\ref{posvel}. Forces acting on the material in the magnetic islands in radial direction are also shown. The pressure gradient and Lorentz forces push the rope radially outwards. 

The launching of the flux rope is caused by the reconnection in the disk, near the disk surface, as shown in the left panel Fig.~\ref{velbs}. In the right panels in the same Figure is shown the reconnection signature in the magnetic field and velocity components perpendicular to the reconnection layer: all three magnetic field components and both poloidal velocity components change sign. Reconnection occurs throughout the disk and in the corona, but its effect on the matter depends on the value of plasma $\beta=P_{\rm gas}/P_{\rm mag}$. Only in the locations where it is about unity, material from the disk will be pushed by reconnection. In the rarefied corona, plasma $\beta$ is much smaller, and in the dense disk, it is much larger than unity.

Inside the disk, which is accreting because of magneto-rotational instability (MRI) providing the sufficient dissipation, reconnection layers which are brought close to the disk surface, can result in the formation of flux rope and its further ejection into the corona. Once lifted into the rarefied corona, the flux rope can be expelled outwards or break. In both cases it would be observed as episodic emission from the vicinity of the black hole.

In addition to the reconnection layer below the magnetic flux rope, there is another reconnection layer, above the magnetic flux rope in our simulations--see Fig.~\ref{velbs}. It helps the opening of the magnetic field lines and ejection of the flux rope.

\section{Conclusions}\label{conclus}
We have performed 3D ideal GRMHD numerical simulations of a hot accretion flow around a black hole, to study formation and motion of flux ropes. During the time-evolution until $t=40000~r_{\mathrm g}/c$, magnetic flux ropes of the azimuthal extension of about $120^\circ$ or less are formed, which show as magnetic islands in 2D slices in colatitudinal planes at different azimuthal angles. These flux ropes are created by reconnection close to the disk surface, where the plasma $\beta$, defined as the ratio of the gas to magnetic pressure, is close to unity.

Because of the reconnection and disk differential rotation, the flux ropes are twisted and pushed radially outwards and launched into the corona, spiralling-out from the central object. The radial velocity of their outward propagation is of the order of 0.01~c.

Ejection of the flux ropes from the disk surface repeats periodically in our simulation, with the period of about $1000~r_{\mathrm g}/c$. It could cause episodic flaring from the vicinity of the disk around a black hole.

In addition to the reconnection layer near the disk surface, which forms the flux rope, another reconnection layer above the flux rope can form, helping its outward launch.

\ack

M\v{C} was supported by CAS President’s International Fellowship for Visiting Scientists (grant No. 2020VMC0002), and the Polish NCN grant 2019/33/B/ST9/01564. FY and HY are supported in part by the National Key Research and Development Program of China (Grant No. 2016YFA0400704), the Natural Science Foundation of China (grants 11633006), and the Key Research Program of Frontier Sciences of CAS (No. QYZDJSSW-SYS008). This work made use of the High Performance Computing Resource in the Core Facility for Advanced Research Computing at Shanghai Astronomical Observatory. We thank the referee for constructive questions and suggestions.
\bibliographystyle{ragtime}
\bibliography{cem}

\end{document}